# Towards Cognitive Load Assessment Using Electrooculography Measures

Arash Abbasi Larki
*Faculty of Electrical Engineering, K. N. Toosi University of Technology,*
Tehran, Iran
arash.abbasilarki@email.kntu.ac.ir

Akram Shojaei
*Faculty of Electrical Engineering, K. N. Toosi University of Technology,*
Tehran, Iran
akram.shojaeibagheini@email.kntu.ac.ir

Mehdi Delrobaei
Department of Mechatronics,
*Faculty of Electrical Engineering, K. N. Toosi University of Technology,*
Tehran, Iran
delrobaei@kntu.ac.ir

*Abstract*— Cognitive load assessment is crucial for understanding human performance in various domains. This study investigates the impact of different task conditions and time constraints on cognitive load using multiple measures, including subjective evaluations, performance metrics, and physiological eye-tracking data. Fifteen participants completed a series of primary and secondary tasks with different time limits. The NASA-TLX questionnaire, reaction time, inverse efficiency score, and eye-related features (blink, saccade, and fixation frequency) were utilized to assess cognitive load. The study results show significant differences in the level of cognitive load required for different tasks and when under time constraints. The study also found that there was a positive correlation ($r = 0.331$, $p = 0.014$) between how often participants blinked their eyes and the level of cognitive load required but a negative correlation ($r = -0.290$, $p = 0.032$) between how often participants made quick eye movements (saccades) and the level of cognitive load required. Additionally, the analysis revealed a significant negative correlation ($r = -0.347$, $p = 0.009$) and ($r = -0.370$, $p = 0.005$) between fixation and saccade frequencies under time constraints.

*Keywords*— Mental workload, EOG, eye movements, eye tracking, gaze tracking, biomechatronic systems.

## I. INTRODUCTION

The cognitive load is crucial in accomplishing everyday tasks and has been extensively researched in psychology, ergonomics [1], and education [2]. The cognitive load can be defined as the mental resources needed to perform a particular task [3]. The concept of cognitive load is related to both the object and the subject involved, meaning that performance can differ between tasks and among individuals performing the same task. In such situations, diverse strategies are employed, and each person selects a distinctive approach to cope with the circumstances. It is also crucial to understand and assess the cognitive load of human operators in high-pressure jobs, such as surgery [4] and aerial navigation [5], since high levels of cognitive load lead to poor performance, risk increase, and low accuracy.

There are no universally recognized tools or techniques for directly measuring cognitive load. However, objective and subjective measures are available for assessing cognitive load [6]. Subjective measurements rely on the operator's perceived feelings and self-evaluations, utilizing questionnaires such as the National Aeronautics and Space Administration-Task Load Index (NASA-TLX) [7] and Subjective Workload Assessment Technique (SWAT) [8] to gauge cognitive load levels.

The NASA-TLX and SWAT are among the multidimensional measures to assess cognitive load. Unidimensional measures like the Rating Scale of Mental Effort (RMSE) are characterized by their simplicity; however, they are limited in their capacity to comprehensively appraise cognitive load as they solely focus on quantifying mental effort without considering other relevant dimensions [9]. While these approaches are convenient to administer, they cannot provide real-time, objective, and precise outcomes.

In contrast, objective measurements predominantly rely on analyzing task performance recordings and physiological signals, which encounter fewer disturbances during the task and alleviate the mentioned limitations. The physiological signals commonly employed in this context can be categorized into several groups, including electroencephalogram (EEG), heart rate, eye movement, respiration, electromyogram (EMG), and skin conductance [10].

Previous studies have shown that specific metrics of eye behavior can provide valuable information about cognitive processes. This information can help recognize such states [11]. Mounting evidence indicates that eye-tracking data is closely related to conventional cognitive evaluation measures. Such evidence suggests that eye-tracking technology can be utilized to assess and keep track of mental states, illness severity, and disease advancement in neurological conditions [12].

Eye-tracking metrics have been identified as promising unobtrusive indices of cognitive load. Eye tracking has many benefits relative to other approaches in that it provides continuous, objective measurements with high temporal and spatial resolutions without burdening or interrupting the user [13]. Eye-tracking is a nonverbal and less cognitively demanding method of assessing cognitive load.

Based on the literature, there are four main methods for eye-tracking [14]: (1) scleral search coil, (2) infrared oculography (IOG), (3) electrooculography (EOG), and (4) video oculography (VOG). Each of these techniques has its advantages and disadvantages.

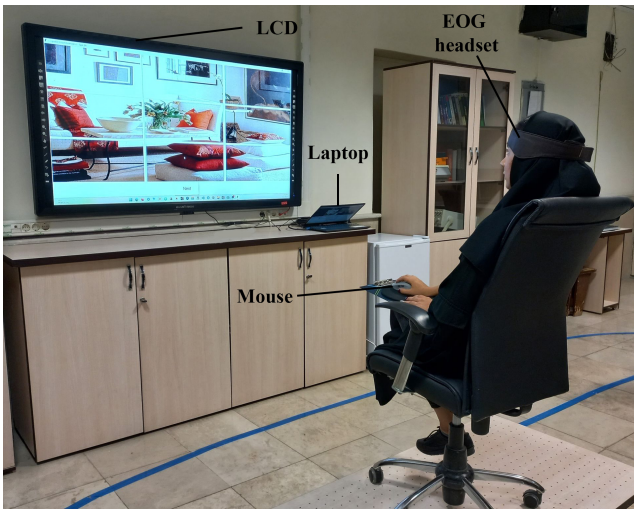

Fig. 1. The experimental setup.

The scleral search coil technique [15] uses a contact lens with mirrors attached to a wire coil, which responds to a magnetic field, generating a signal representing eye position. The technique enables the detection of coil orientation in a magnetic field, allowing accurate recording of eye movements by monitoring the infrared wavelength range reflected by the mirror.

Infrared oculography (IOG) [16] is a technique that measures the infrared light strength mirrored from the sclera to gather information about eye position. The method uses a pair of glasses to emit light and relies on light and pupil detection algorithms. A reference point called a corneal reflection or glint is introduced to address head movement sensitivity using an infrared light source.

Electrooculography (EOG) is a highly advantageous and cost-effective human-computer interaction (HCI) method [17]. This non-invasive and not complicated approach involves attaching sensors around the eyes to detect the electric field generated during eye rotations, utilizing skin fluctuations. It enables separate recording of horizontal and vertical eye movements through electrodes. Additionally, EOG is suitable for implementing machine learning algorithms and real-time processing due to its high temporal and spatial resolution. Moreover, EOG systems exhibit low power consumption, making them efficient for prolonged usage.

The video oculography (VOG) [18] technique involves a typical setup consisting of a camera recording eye movements and a computer analyzing the gaze data. It can use visible or infrared lights and is a non-invasive remote eye-tracking system. The VOG can be implemented with either a single camera or multiple cameras. Single-camera systems use infrared light to create a corneal reflection as a reference point for gaze estimation. However, they have a limited field of view. To address head movement, remote VOG trackers may use two stereo cameras or one wide-angle camera to locate the person and zoom in on their face. Multiple camera eye trackers achieve a larger field of view for free head motion, using one camera for the eye and another for capturing head location. The data from these cameras are combined to estimate the point of gaze, often utilizing a stereo-vision system for calibration [14].

The swift advancement of wearable devices has facilitated the acquisition of diverse physiological data from human operators in a non-intrusive manner. Additionally, these devices can perform tasks without interrupting the primary task [19]. This progress has rendered the collection of such data more practical and viable.

In the proposed work, participants performed a visual search task as the primary task along with a challenging working memory task as the secondary task. A wearable EOG headset was used to record their eye movements, and after each task, they filled out a NASA-TLX questionnaire. Subsequently, the outcomes derived from the eye-tracking device and the self-report questionnaires were analyzed to investigate potential correlations between eye-related features and cognitive load levels.

The paper is organized as follows: Section II describes the study design, experimental setup, and signal processing methods. Section III displays the evaluation results and subsequent discussions, while Section IV concludes the paper.

## II. METHODS

### A. Participants

Seventeen individuals, primarily undergraduate students, participated in the experiments. However, two were excluded from the study because insufficient data was recorded (this was part of the exclusion criteria). The data analysis was conducted with information gathered from the remaining fifteen individuals (five identified as female), with an average age of 22.3 years.

### B. Experimental setup

We chose twenty indoor environment images from a publicly available dataset [13]. We divided each image into nine equal squares for our experiments. The participants were instructed to select the squares with a particular object as their primary task. In the following stages of the experiment, they were given an additional task of counting backward from 1000 in intervals of 7, alongside their primary task.

The images were displayed on a 46-inch LCD screen. The participants were positioned in a way that their eyes were precisely in the center of the screen in a horizontal line, as shown in Fig. 1. To capture eye activities, a two-channel EOG headset with five electrodes was utilized to record the vertical and horizontal movements (Fig. 2). The placement of the vertical, horizontal, and reference electrodes is shown in Fig. 3. The data were obtained at a sampling rate of 250 Hz with a 4-bit resolution, then transmitted to a PC via Bluetooth. An application was designed in MATLAB to visualize and store the recordings, which displayed the EOG signal in real-time. A graphical user interface (GUI) in Python was also developed to display images and record the time and position of the clicks to conduct the primary task. To initialize the experiment, each participant was randomly assigned four sets of twenty images, each containing five images.

### C. Procedure

At the start of the study, all participants signed an informed consent form. Each participant was then seated on an adjustable chair, positioning their eyes at 160 (cm) from the display. During data collection, the participants were instructed to maintain stability in their head movements. Before the experiment began, participants were given instructions and shown a sample image to become familiar with the procedure as shown in Fig. 4.

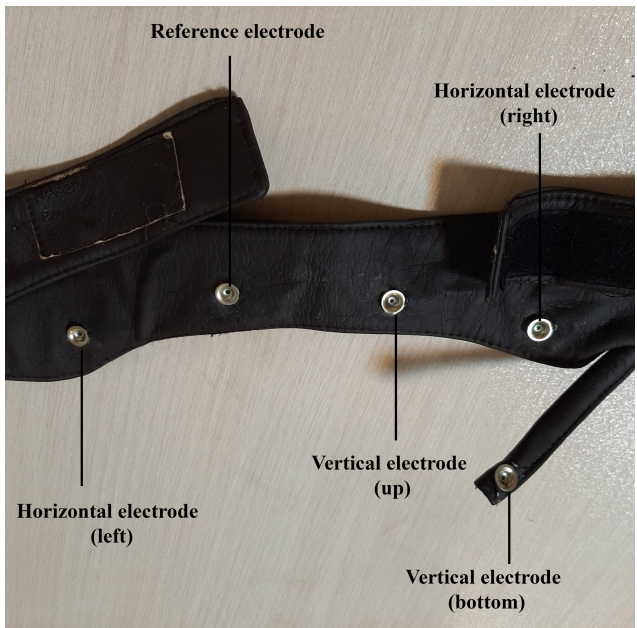

Fig. 2. The EOG headset used in this study; with two channels and five electrodes to record eye activity, such as vertical and horizontal movements and blinking.

The study procedure was segmented into four parts: the primary task (no time constraint), the primary task with a time constraint, the primary task with a secondary task (no time constraint), and the primary task with both a secondary task and a time constraint. In the first part, we asked the participant to complete the primary task without time constraints. In the second part, the subject was supposed to complete the primary task as fast as possible. In the third part, the participant was supposed to complete the primary and secondary tasks simultaneously at a comfortable pace. The last part was the same as the third part but with time constraints. The primary task in each part consisted of 5 distinct indoor images selected randomly. The subject was asked to complete all the parts, fill out the self-report form (NASA-TLX) after each part, and rest for 2 minutes between every two consecutive parts.

### D. Cognitive load assessment

There exist four distinct approaches for cognitive load assessment, namely subjective measures, performance measures, psychophysiological measures, and analytical measures. In our study, we employed the first three methodologies to assess the level of cognitive load. The primary task served as the basis for evaluating cognitive load. In contrast, the secondary task was deliberately chosen to induce cognitive load among the participants during the execution of the third and fourth phases.

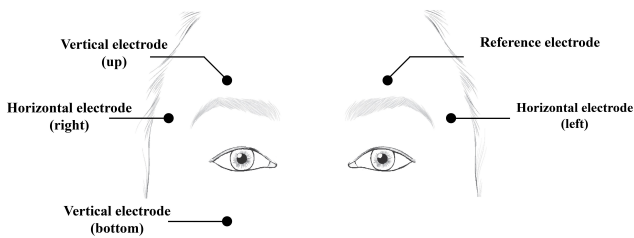

Fig. 3. The placement of the electrodes on the head.

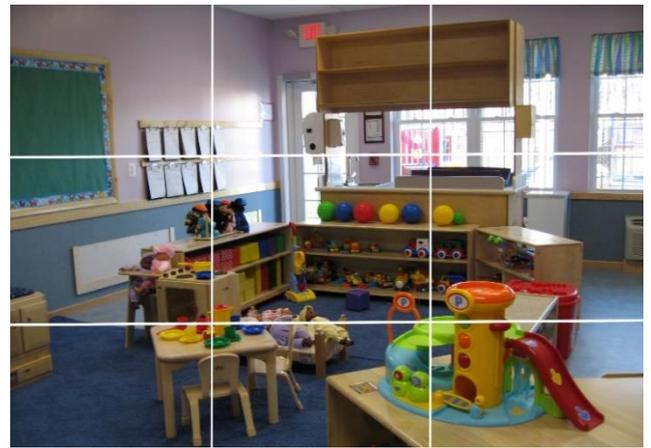

Fig. 4. A sample test image shown to the participants before performing task 1.

#### 1) Subjective measurement

The original NASA-TLX survey consisted of two main parts, where the total load was assessed through six subjective subscales. These subscales were presented on a single page, forming one part of the questionnaire: mental demand, physical demand, temporal demand, performance, effort, and frustration.

For each of these subscales, there was a corresponding description that the participants needed to read before providing their ratings. The ratings were given for each task using a scale of 100 points with 5-point intervals. The individual ratings for each subscale were then combined to calculate the task load index. Participants could provide more accurate and meaningful responses by offering detailed descriptions for each measurement.

#### 2) Performance measure

Throughout all phases of the study, researchers recorded the count of errors made and correct squares missed. They also measured the time taken to complete each phase, known as reaction time (RT). A commonly used metric, inverse efficiency score (IES), was employed to integrate speed and accuracy. The IES is computed by dividing the mean RT by each participant's percentage of correct answers (PC) in a specific condition. PC was calculated by dividing the number of correct answers by the sum of the correct, incorrect, and missed answers.

#### 3) Physiological measure

EOG signal was pre-processed and was used to extract eye-related features, namely blink frequency, saccade frequency, and fixation frequency.

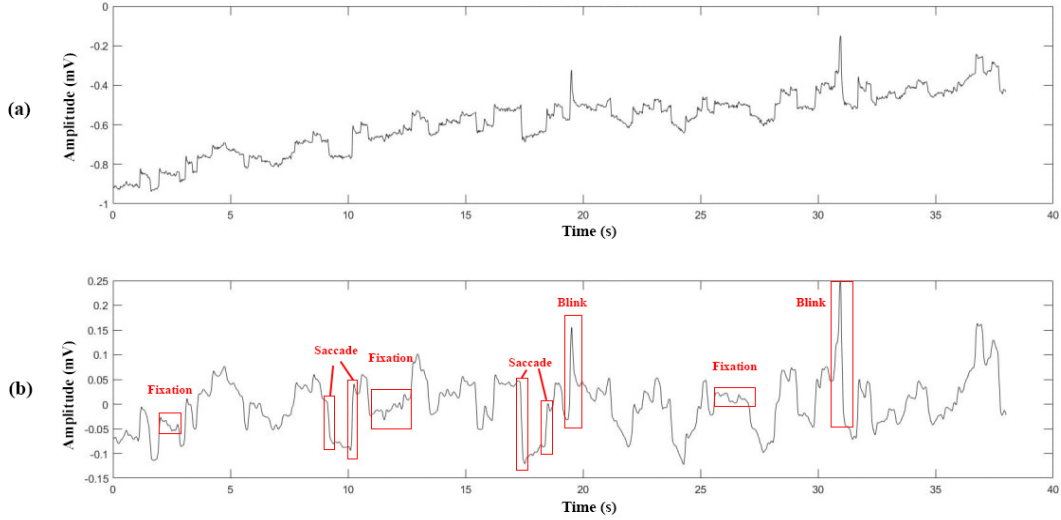

Fig. 5. A sample of the recorded EOG signals. (a) The raw signal, (b) the preprocessed signal with feature labels.

*E. Data analysis methodology*

Due to the nature of the wearable sensors, the signal inevitably contains noise and artifacts, so some preprocessing techniques needed to be implemented to remove undesirables.

*1) Raw data pre-processing*

Several key functions were employed in the data processing workflow to enhance the quality and extract meaningful patterns from the signal. The initial step involved applying the moving median filter on the raw signal, denoted as $S_{raw}$. This filter leveraged a sliding window of size $w_{size}$ to compute the median value of each data point, leading to smoothed datasets denoted as $S_{smth}$. Mathematically, the moving median $m_i$ at data point $s_i$ can be expressed as:

$$k = {w_{size}}/{2} \qquad (1)$$

$$m_i = median(\{s_{i-k}, \ldots, s_{i+k-1}, s_i, s_{i+1}, \ldots, s_{i+k}\}) \qquad (2)$$

Subsequently, detrending was performed to remove any long-term trends from the original datasets, resulting in detrended datasets $S_{det}$. Detrending was achieved by subtracting the corresponding smoothed data $S_{smth}$ from the raw dataset $S_{raw}$.

$$S_{det} = S_{raw} - S_{smth} \qquad (3)$$

For the final stage of data processing, a finite impulse response (FIR) filter was applied to the detrended dataset $S_{det}$, using the filter coefficients $c$ with a uniform window length. The filtering process can be expressed mathematically as follows:

$$S_{filt_i} = \sum_j c_j \cdot d_{i-j} \qquad (4)$$

where $S_{filt_i}$ represents the filtered value at index $i$ in the dataset $S_{filt}$, $c_j$ denotes the $j$-th coefficient of the FIR filter, and $d_{i-j}$ denotes the data point at index $i-j$ in the detrended dataset $S_{det}$. This FIR filtering step efficiently minimized noise and further accentuated the underlying patterns in the data, yielding a refined and informative dataset $S_{filt}$.

*1) Event detection*

Eye-related events in the electrooculography signal were identified using human coders [20]. Additionally, a human expert verified the assigned labels for the detected signal events to enhance accuracy and reliability. Fig. displays instances of the raw signal, pre-processed signal, and selected samples of blinks, saccades, and fixations for illustration purposes.

## III. RESULTS AND DISCUSSION

The mean values for each subscale and the corresponding mean scores across all activities are displayed in Table 1. An investigation was conducted to examine the cognitive load of the four tasks in the experiment. To achieve this, a one-way repeated measures ANOVA was utilized on the NASA-TLX reports submitted by the participants. The findings revealed significant statistical variations among the four tasks ($p < .05$), implying an escalation in cognitive load as the experiment progressed from task 1 to task 4. Table 2 provides the results of statistical analyses investigating the impact of different task conditions (multi-task vs. single-task) and time constraint (with time constraint vs. without time constraint) on various measures (mental demand, physical demand, temporal demand, performance, effort, frustration, reaction time, IES, fixation frequency, saccade frequency, and blink frequency) using two-way repeated measures ANOVA. The results indicate that task conditions significantly influenced several measures. Specifically, mental demand, physical demand, temporal demand, performance, effort, frustration, reaction time, IES, and blink frequency all exhibited significant differences ($p < .05$) between multi-task and single-task scenarios. This suggests that the nature of a task, whether involving multiple tasks or a single task, substantially impacts these measures. Regarding the effect of time constraint, mental demand, temporal demand, effort, and reaction time, IES, and saccade frequency showed significant differences ($p < .05$) between tasks with and without time constraints. These findings confirm the increase in cognitive load as the experiment continues from Task 1 to Task 4. We calculated the Pearson correlation coefficients and assessed their statistical significance to investigate the relationships between blink, fixation, and saccade frequency with cognitive load levels ($p < .05$). The findings indicated significant positive correlations between blink frequency and cognitive load level ($r = 0.331$, $p = .014$), suggesting that higher blink frequencies are associated with increased cognitive load level. Additionally, a negative correlation was observed between time constraint and fixation frequency ($r = -0.347$, $p = .009$), indicating that time constraint leads to a rise in fixation frequency. Moreover, higher cognitive load levels negatively correlated with saccade frequency ($r = -0.290$, $p = .032$). Lastly, a negative correlation was identified between saccade frequency and time constraint ($r = -0.370$, $p = .005$), indicating a connection between time constraint and higher saccade frequency.

Table 1. Mean values (±SD) of NASA-TLX scores across all tasks

| Measure | Task 1 | Task 2 | Task 3 | Task 4 |
|---|---|---|---|---|
| Mental demand | 23.55 (±17.35) | 27.30 (±12.6) | 66.65 (±20.75) | 76.3 (±20.55) |
| Physical demand | 13.90 (±15.20) | 11.00 (±9.10) | 25.00 (±26.75) | 32.00 (±32.85) |
| Temporal demand | 23.90 (±18.70) | 33.30 (±25.25) | 42.00 (±31.2) | 69.30 (±27.40) |
| Performance | 83.55 (±21.85) | 85.00 (±14.35) | 50.65 (±29.45) | 52.30 (±22.90) |
| Effort | 23.70 (±21.05) | 35.30 (±25.30) | 68.3 (±17.95) | 72.65 (±20.90) |
| Frustration | 16.05 (±12.40) | 18.65 (±17.05) | 47.00 (±27.95) | 54.30 (±31.00) |
| Mean | 30.78 (±23.93) | 35.09 (±23.82) | 49.93 (±13.68) | 59.48 (±14.07) |

Table 2. Two-way repeated measures ANOVA results of different task conditions.

| Measure | Task | | Time | | Int.[a] |
|---|---|---|---|---|---|
| | *p-value* | *Dif. (single-multi)* | *p-value* | *Dif. (without-with)* | *p-value* |
| Mental demand | **<.001** | -44.500 | **.039** | -6.167 | .380 |
| Physical demand | **.030** | -15.500 | .519 | -2.167 | **.028** |
| Temporal demand | **<.001** | -26.167 | **.005** | -18.167 | .109 |
| Performance | **<.001** | 34.000 | .643 | -1.667 | .815 |
| Effort | **<.001** | -40.833 | **.032** | -7.500 | .175 |
| Frustration | **<.001** | -34.667 | .298 | -4.333 | .689 |
| Reaction time | **.008** | -9.400 | **<.001** | 13.600 | **.017** |
| IES | **<.001** | -48.053 | **<.001** | 47.989 | **.031** |
| Fixation frequency | **.008** | 0.288 | .814 | -0.018 | .643 |
| Saccade frequency | .111 | 0.183 | **.004** | 0.265 | .183 |
| Blink frequency | **.005** | -0.102 | .075 | 0.033 | .859 |

[a] Interaction

## IV. CONCLUSION

The cognitive load was assessed using subjective measurements (NASA-TLX), performance measures (RT and IES), and physiological measures (eye-related features such as blink frequency, saccade frequency, and fixation duration). The statistical analysis, employing two-way repeated measures ANOVA, revealed significant differences in several measures based on task conditions and time constraints. Specifically, mental demand, physical demand, temporal demand, performance, effort, frustration, reaction time, IES, saccade frequency, and blink frequency demonstrated significant variations between multi-task and single-task scenarios. Additionally, temporal demand, RT, IES, fixation frequency, and saccade frequency showed significant differences between tasks with and without time constraints. The results highlight the importance of considering task conditions and time constraints in assessing cognitive load and suggest that eye-tracking measures, especially EOG, can provide valuable insights into cognitive load levels.

This research contributes to understanding cognitive load assessment methods and their potential applications in various fields where task performance and human factors are crucial considerations. Further studies could explore additional physiological and performance measures to deepen our understanding of cognitive load dynamics and develop more robust and precise assessment approaches.

REFERENCES

[1] M. S. Young, K. A. Brookhuis, C. D. Wickens, and P. A. Hancock, "State of science: mental workload in ergonomics," Ergonomics, vol. 58, no. 1, pp. 1–17, Dec. 2014.

[2] K. J. Wilby and B. Paravattil, "Cognitive load theory: Implications for assessment in pharmacy education," *Research in Social and Administrative Pharmacy*, Dec. 2020.

[3] M. S. Young, K. A. Brookhuis, C. D. Wickens, and P. A. Hancock, "State of science: mental workload in ergonomics," *Ergonomics*, vol. 58, no. 1, pp. 1–17, Dec. 2014.

[4] M. Inama et al., "Cognitive load in 3d and 2d minimally invasive colorectal surgery," *Surgical Endoscopy and Other Interventional Techniques*, vol. 34, no. 7, pp. 3262–3269, Apr. 2020.

[5] P. A. Hebbar, K. Bhattacharya, G. Prabhakar, A. A. Pashilkar, and P. Biswas, "Correlation Between Physiological and Performance-Based Metrics to Estimate Pilots' Cognitive Workload," *Frontiers in Psychology*, vol. 12, Apr. 2021.

[6] J. A. Blanco et al., "Quantifying Cognitive Workload in Simulated Flight Using Passive, Dry EEG Measurements," *IEEE Transactions on Cognitive and Developmental Systems*, vol. 10, no. 2, pp. 373–383, Jun. 2018.

[7] J. Heard, C. E. Harriott, and J. A. Adams, "A survey of workload assessment algorithms," *IEEE Transactions on Human-Machine Systems*, vol. 48, no. 5, pp. 434–451, Oct. 2018.

[8] S. G. Hart and L. E. Staveland, "Development of NASA-TLX (task load index): Results of empirical and theoretical research," *Advances in Psychology*, vol. 52, pp. 139–183, 1988.

[9] G. B. Reid and T. E. Nygren, "The Subjective Workload Assessment Technique: A Scaling Procedure for Measuring Mental Workload," *Advances in Psychology*, pp. 185–218, 1988.

[10] S. Frazier, B. J. Pitts, and S. McComb, "Measuring cognitive workload in automated knowledge work environments: a systematic literature review," *Cognition, Technology & Work*, vol. 24, no. 4, pp. 557–587, Aug. 2022.

[11] K. Guan, Z. Zhang, T. Liu, and H. Niu, "Cross-Task Mental Workload Recognition Based on EEG Tensor Representation and Transfer Learning," *IEEE Transactions on Neural Systems and Rehabilitation Engineering*, vol. 31, pp. 2632–2639, Jan. 2023.

[12] V. Skaramagkas et al., "Review of eye tracking metrics involved in emotional and cognitive processes," *IEEE Reviews in Biomedical Engineering*, pp. 1–1, 2021.

[13] E. Ktistakis et al., "COLET: A dataset for COgnitive workLoad estimation based on eye-tracking," *Computer Methods and Programs in Biomedicine*, vol. 224, p. 106989, Sep. 2022.

[14] R. Mallick, D. Slayback, J. Touryan, A. J. Ries, and B. J. Lance, "The use of eye metrics to index cognitive workload in video

games," *Zenodo (CERN European Organization for Nuclear Research)*, Oct. 2016.

[15] F. Klaib, N. O. Alsrehin, W. Y. Melhem, H. O. Bashtawi, and A. A. Magableh, "Eye tracking algorithms, techniques, tools, and applications with an emphasis on machine learning and Internet of Things technologies," *Expert Systems with Applications*, vol. 166, p. 114037, Mar. 2021.

[16] C. R. Picanço and F. Tonneau, "A low-cost platform for eye-tracking research: Using Pupil© in behavior analysis," *Journal of the Experimental Analysis of Behavior*, vol. 110, no. 2, pp. 157–170, Jun. 2018.

[17] S. Milanizadeh and J. Safaie, "EOG-Based HCI System for Quadcopter Navigation," *IEEE Transactions on Instrumentation and Measurement*, vol. 69, no. 11, pp. 8992–8999, Nov. 2020.

[18] J. Larrazabal, C. E. García Cena, and C. E. Martínez, "Video-oculography eye tracking towards clinical applications: A review," *Computers in Biology and Medicine*, vol. 108, pp. 57–66, May 2019.

[19] R. L. Charles and J. Nixon, "Measuring mental workload using physiological measures: A systematic review," *Applied Ergonomics*, vol. 74, pp. 221–232, Jan. 2019.

[20] Birtukan Birawo and Pawel Kasprowski, "Review and Evaluation of Eye Movement Event Detection Algorithms," *Sensors*, vol. 22, no. 22, pp. 8810–8810, Nov. 2022.